\documentclass{moriond}

\def\Journal#1#2#3#4{{#1} {\bf #2}, #3 (#4)}

\def\PRD{{\em Phys. Rev.} D}

\def\be{\begin{equation}}
\def\ee{\end{equation}}
\def\bea{\begin{eqnarray}}
\def\eea{\end{eqnarray}}

\newcommand{\Photo}{\includegraphics[height=35mm]{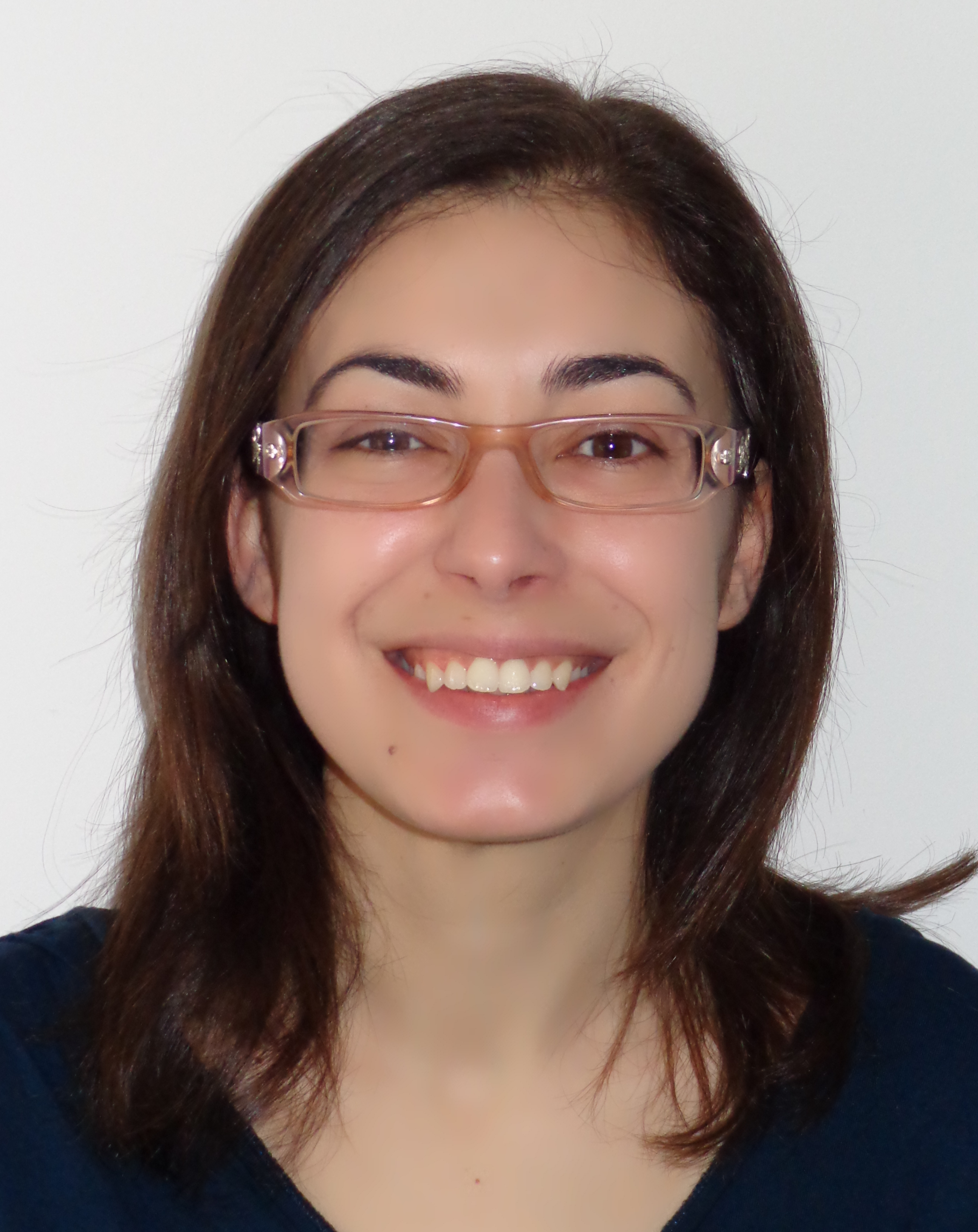}}

\begin{document}
\vspace*{4cm}
\title{DARK MATTER PRODUCED IN ASSOCIATION WITH TOP QUARK PAIR}

\author{ D. PINNA\\
on behalf of the CMS Collaboration}
\address{Department of Physics, University of Zurich, 190 Winterthurerstrasse,\\
8057 Zurich, Switzerland}

\maketitle\abstracts{
A search for dark matter produced in association with a top quark pair is presented. The search is performed using $19.7 \mathrm{fb^{-1}}$ of proton-proton collisions recorded at a center of mass energy of 8 TeV with the CMS detector at the LHC. The signature investigated is top quark pairs in the semi-leptonic final state plus missing transverse energy. This work focuses in particular on dark matter production through scalar interaction where a proportionality to the quark mass is expected. }

\section{Introduction}
Astrophysical studies provide precise information about the presence of dark matter (DM) in our universe and its abundance~\cite{ko}. These observations cannot be included into the standard model (SM), which is able to describe only the visible matter, which makes up 4\% of the mass of the universe. Dark matter instead accounts for about 24\% of the mass of the universe, but so far no information about its nature nor non-gravitational interactions is available.\\
In DM searches, a common assumption is to interpret DM as a weakly interacting massive particles, which interacts with SM particles. Under this hypothesis DM particles can be detected through dedicated experiments~\cite{xe,pa}, and it could be produced in proton-proton collisions at LHC. Collider data provides access to a different range of possible interactions between SM and DM particles with respect to current dedicated experiments, allowing an important interplay and complementarity among experiments to discover DM.\\
Many specific models provide a DM particle candidate in agreement with astrophysical observations~\cite{be,fe}, making essential a model-independent DM searches. A possible approach is given by an effective field theory, where the interaction between DM and SM particles is parameterized by effective operators.
\subsection{Effective field theory and scalar operator}\label{subsec:prod}
\begin{figure}
\begin{minipage}{0.33\linewidth}
\end{minipage}
\hfill
\begin{minipage}{0.32\linewidth}
\centerline{\includegraphics[width=0.6\linewidth]{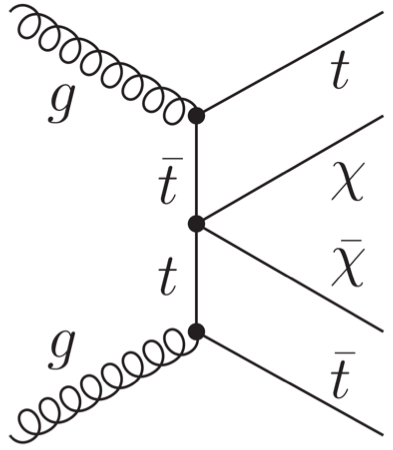}}
\end{minipage}
\hfill
\begin{minipage}{0.32\linewidth}
\end{minipage}
\caption[]{Dominant diagram contributing to the production of DM particles in association with top quarks at the LHC.}
\label{fig:feyn}
\end{figure}

Effective field theory (EFT) approach is valid only when the energy of the interaction is such that the details of the mediators are not resolved~\cite{bu}. To avoid this limitation the particle that mediates the interaction of the DM particles with the particles of the SM can be assumed to be somewhat heavier than the DM particle itself. Under this assumption, the interaction can be seen as between the DM and the SM particles directly and parameterized by effective operators~\cite{go}.\\
Assuming a scalar interaction between the DM and the SM particles, the strength of the interaction is proportional to a Yukawa term. For example, assuming that the DM particle is a Dirac fermion $\chi$, the effective scalar operator can be expressed as~\cite{bel}:
\be
L_{int}=\frac{m_{q}}{M_{*}^{3}}q\bar{q}\chi\bar{\chi}
\label{eq:scalOp}
\ee
where $m_{q}$ is the mass of the quark $q$ that interacts with the DM particle and $M_{*}$ is the interaction scale.\\
As a consequence, couplings to light quarks are suppressed and the sensitivity to the scalar interaction can be improved by searching for DM particles in final states with third-generation quarks~\cite{ton}. This motivates the search for the production of DM particles in association with a pair of top quarks through a scalar interaction. In this work such study is done using the data collected by the CMS experiment~\cite{cms} at LHC~\cite{b2g}.\\
The dominant associated Feynman diagrams for this process is given by the interaction of gluons from the protons, a top quark pair together with two DM particles, as shown in Fig.~\ref{fig:feyn}. The top quark decays into a W boson and a b quark and the semileptonic final state, where one W boson decays leptonically and the other one hadronically, is studied in the analysis performed at $\sqrt{s} = 8$ TeV. The advantages of this final state are coming from the high branching ratio combined with a clean signature.
\section{Analysis strategy}
Signal events will be expected to contain large missing transverse energy (MET) in their final state, as well as one lepton (electron or muon) and four jets, two of which originate from b quarks. The optimal selection for the signal topology is obtained requiring one lepton and at least three jets of which at least one coming from a b quark. SM processes, referred as background processes, can provide the same signature. Relevant backgrounds for this search include $\mathrm{t\bar{t}}$, single top and $\mathrm{W+jets}$, Di-bosons, and Drell-Yan events. To improve the sensitivity of the analysis it is essential to reject background events and precisely determine the remaining contributions. \\
Background contaminations can be reduced using kinematical differences with the signal.
Using simulation, the DM signal shows a larger MET than the backgrounds because of the two DM particles escaping the detector.\\
After a minimal requirement on the MET, the main background contributions originate from $\mathrm{W+jets}$ and $\mathrm{t\bar{t}}$ events containing a single leptonically decaying W boson. The transverse mass $M_{T} = \sqrt{2p_{lep}E_{miss}(1 − \cos(\Delta\phi))}$,  where $p_{l}$ is the transverse momentum of the lepton and $\Delta\phi$ is the opening angle in azimuth between the lepton and MET vector, is constrained kinematically to $M_{T} < M_{W}$ for the on-shell W boson decay in the $\mathrm{t\bar{t}}$ and $\mathrm{W+jets}$ events.\\
The two leading jets ($j_{1}$ and $j_{2}$) and the MET tend to be more separated in $\phi$ in signal events than in $\mathrm{t\bar{t}}$ events, therefore a selection on this variable helps to further enhance the signal sensitivity.\\
The most challenging background for this analysis comes from $\mathrm{t\bar{t}}$ events where one of the leptons is unobserved, leading to high values of MET. To distinguish signal from background events, the mass of the particle from which the MET originates can be estimated using kinematical constraints. This is achieved using the $M_{T2}^{W}$ variable~\cite{mt2w}, which is constrained to the top quark mass for $\mathrm{t\bar{t}}$ events and it will have higher values for signal events.\\
The selections found to be optimal to define the signal region (SR) for this analysis are: MET$> 320$ GeV, $M_{T} > 160$ GeV, $min\Delta\phi(j_{1,2},MET) > 1.2$, and $M_{T2}^{W} >200$ GeV.\\
\begin{figure}
\hfill
\begin{minipage}{0.33\linewidth}
\centerline{\includegraphics[width=1.05\linewidth]{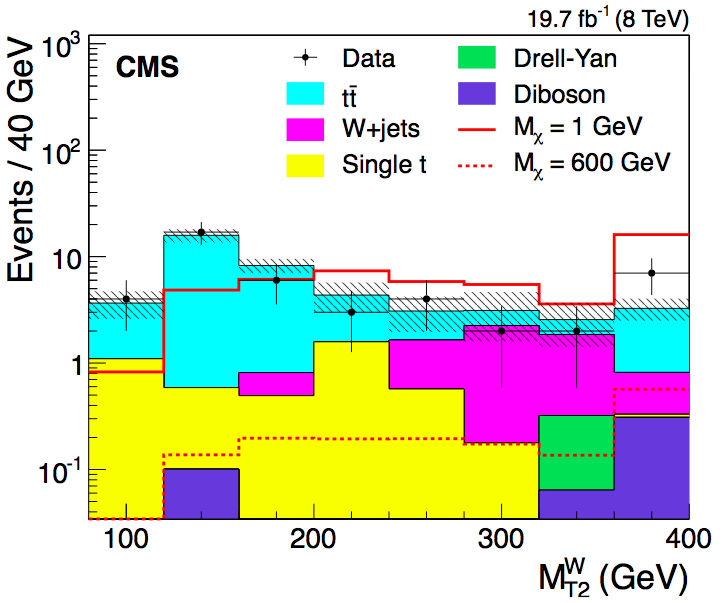}}
\end{minipage}
\hfill
\begin{minipage}{0.32\linewidth}
\centerline{\includegraphics[width=1.05\linewidth]{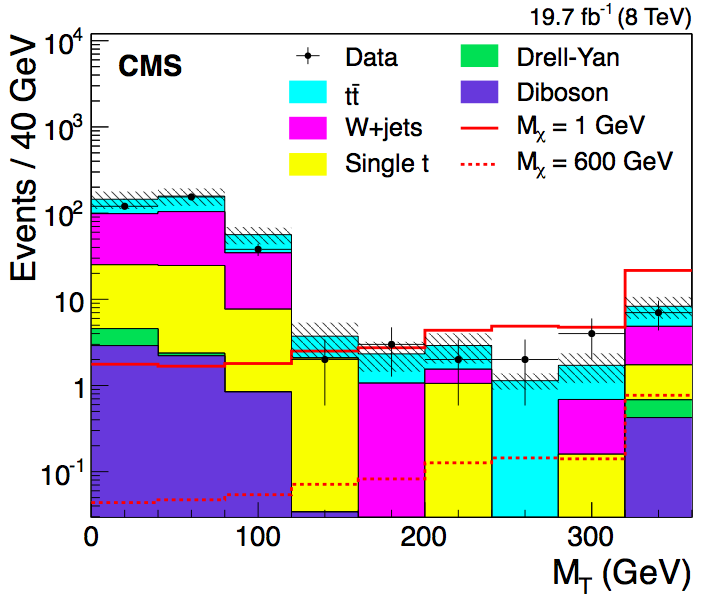}}
\end{minipage}
\hfill
\hfill
\caption[]{Distributions of $M{T}$ (left) and $M_{T2}^{W}$ (right), each plotted after applying all other selections, showing the discriminating power between signal and background. Two simulated DM signals with mass $M_{\chi}$ of 1 and 600 GeV and an interaction scale $M_{*}$ of 100 GeV are included for comparison. The hatched region represents the total uncertainty in the background prediction.}
\label{fig:radish}
\end{figure}
The normalization for the $\mathrm{t\bar{t}}$ and $\mathrm{W+jets}$ simulated distributions is determined from data to achieve higher precision estimates. The other remaining SM processes contribute around 20\% to the total backgrounds and are taken directly from simulation.\\
The normalization estimation from data consists in evaluating the background yields in a control region (CR). The predicted background yields and their uncertainties are then extrapolated from the CR to the SR using the shape of the distributions from simulation. In this analysis, two CRs have been used, one enriched in $\mathrm{W+jets}$ events and another enriched in $\mathrm{t\bar{t}}$ events~\cite{b2g}. This technique helps to further improve the precision of the background estimation, constraining the systematic uncertainties affecting the shape and the normalization. A total background uncertainty of about 13\% is obtained.

\section{Results}
\begin{figure}
\begin{minipage}{0.33\linewidth}
\end{minipage}
\hfill
\begin{minipage}{0.32\linewidth}
\centerline{\includegraphics[width=1.35\linewidth]{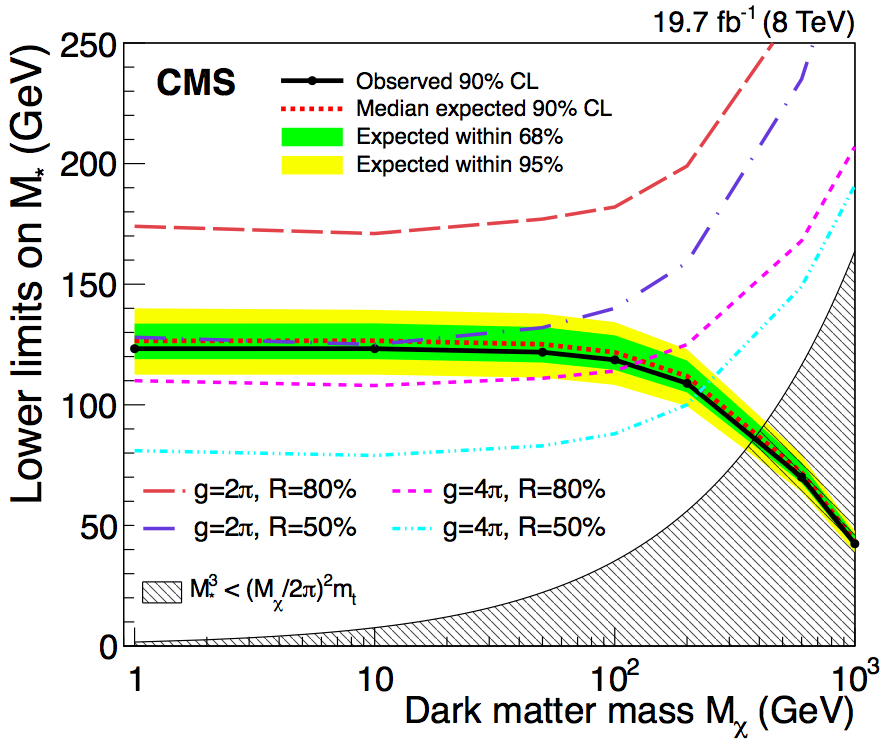}}
\end{minipage}
\hfill
\begin{minipage}{0.32\linewidth}
\end{minipage}
\caption[]{Observed and expected lower limits at 90\% CL on scalar interactions. A lower bound on the validity of the EFT is given the grey-hatched area. The four curves correspond to lower limits on the results for the situations were a fraction R of 50\% and 80\% of the events satisfies the momentum transfer conditions for $g=4\pi$ and $g=2\pi$.}
\label{fig:limit}
\end{figure}

The number of events observed in the SR are presented in Table~\ref{tab:yield}, as well as the expected number of background and signal events for a DM particle with mass of $M_{\chi} = 1$ GeV and an interaction scale $M_{*} = 100$ GeV. No excess of events in the SR is observed and 90\% confidence level (CL) upper limits can be set on the production cross section of DM particles in association with a pair of top quarks, as shown in Fig.~\ref{fig:limit}. 
\begin{table}[t]
\caption[]{Expected number of background events in the SR, expected number of signal events for a DM particle with the mass $M_{\chi}= 1$ GeV, assuming an interaction scale $M_{*}= 100$ GeV, and observed data. The statistical and systematic uncertainties are given on the expected yields.}
\label{tab:yield}
\vspace{0.3cm}
\begin{center}
\begin{tabular}{c|c}
\hline
Source & Yield ($\pm$stat $\pm$syst) \\ \hline
$t\bar{t}$ &  $8.2\pm0.6\pm1.9$ \\
W & $5.2\pm1.8\pm2.1$ \\
Single top &  $2.3\pm1.1\pm1.1$ \\
Diboson & $0.5\pm0.2\pm0.2$ \\
Drell--Yan & $0.3\pm0.3\pm0.1$ \\ \hline
Total Bkg & $16.4\pm2.2\pm2.9$ \\
Signal & $38.3\pm0.7\pm2.1$ \\
Data & 18  \\ \hline
\end{tabular}
\end{center}
\end{table}
Assuming a DM particle with a mass of 100 GeV, interaction scales at 90\% CL below 118 GeV can be excluded. This represents the best limit up to date on the scalar interaction between SM and DM particles performed at CMS.\\
In this analysis, DM production is modeled by an EFT. The couplings g between the mediator and the DM-SM particles should be below the perturbative regime for the EFT to be valid and the momentum transfer $Q_{tr}$ in the event has to be small compared to the mediator mass~\cite{bu}. The region of parameter space in Fig.~\ref{fig:limit} that does not meet the perturbative condition is indicated by the grey-dashed area.  The momentum transfer requirements is considered showing the lower limits on the results for the cases were a fraction R of 50\% and 80\% of the events satisfies the momentum transfer conditions for $g = 4\pi$ and $g = 2\pi$. 

\end{document}